\documentclass[aps,nofootinbib,twocolumn]{revtex4}
\def\be{\begin{equation}}
\def\ee{\end{equation}}
\def\ba{\begin{eqnarray}}
\def\ea{\end{eqnarray}}

\usepackage{amsmath,bbm}
\usepackage{amssymb}
\usepackage{mathrsfs}
\usepackage{mathtools}
\usepackage{amsthm}
\usepackage{xcolor}
\usepackage{verbatim}
\usepackage[pdftex]{graphicx}
\usepackage{hyperref}
\hypersetup{
	colorlinks=true,
	linkcolor=blue,
	filecolor=magenta,      
	urlcolor=cyan,
}

\DeclareFontFamily{U}{rsfs}{}         
\DeclareFontShape{U}{rsfs}{m}{n}{<5> rsfs5 <6><7> rsfs7          %
  <8><9><10><10.95><12><14.4><17.28><20.74><24.88> rsfs10}{}     %
\DeclareMathAlphabet{\mathfs}{U}{rsfs}{m}{n}                     %
\newcommand{\mfs}[1]{\mathfs {#1}}                               %
\newcommand{\n}{{\nonumber}}

\newcommand{\sI}{{\mfs I}}

 \usepackage{braket}

\newcommand{\RN}[1]{%
  \textup{\uppercase\expandafter{\romannumeral#1}}%
}
\newcommand{\out}{\text{out}}

\newcommand{\van}{\scriptstyle}
\usepackage{braket}

\begin{document}

\title{Black Hole Entropy and Planckian Discreteness}

\author{Alejandro Perez}
\email{perez@cpt.univ-mrs.fr}
\affiliation{Aix Marseille Univ, Universit\'e de Toulon, CNRS, CPT, 13000 Marseille, France}

\begin{abstract}
A brief overview of the discovery that macroscopic black holes are thermodynamical systems is presented. They satisfy the laws of thermodynamics and are associated with a temperature and an entropy equal to one quarter of their horizon area in Planck units. They emit black body radiation and slowly evaporate as a consequence of Heisenberg's uncertainty principle.
The problem of understanding the microscopic source of their large entropy, as well as the nature of their final fate after evaporation, are discussed from the perspective of approaches to quantum gravity that predict discreteness at the Planck scale. We review encouraging first steps in computing black hole entropy and briefly discuss their implications for the black hole information puzzle.
\end{abstract}

\maketitle

\section{Introduction}

Black holes are predicted by general relativity as a consequence of gravitational collapse. 
If left in isolation, the highly dynamical phase of formation, is followed by one where the black hole settles down to a (time independent) stationary state
as energy is emitted out to infinity in the form of matter and gravitational radiation. The situation 
is described by  the Kerr-Newman family of stationary solutions of Einsteins equations \cite{wald}.  
Such is the physical picture proposed by the no-hair conjecture (see \cite{wald} for precise account and references). 
According to this picture the memory of the initial conditions that led to the black hole formation seem completely
lost to the outside observers (at least in the classical description).
As for everyday systems in thermal equilibrium, like a box filled with a vast number of gas particles where the initial conditions of individual molecules 
are lost in the thermodynamic description, such property of black holes is the first indication that they too are thermodynamical systems.  

Kerr-Newman black holes are labelled by their mass $M$, angular momentum $J$, and electric charge $Q$.
When  brought out of equilibrium (for instance by sending some matter or radiation into the black hole)  
they return to equilibrium to a different state.  The initial and final equilibrium states are related by the first law  \cite{Bardeen:1973gs}
\be\label{first}
\delta M=\frac{\kappa}{8\pi} {\delta A}+\Omega \delta J+\Phi \delta Q,
\ee 
where $A$ represents the area of the black hole horizon, the surface gravity $\kappa$ is a measure of the gravitational acceleration at the horizon when redshifted to infinity, 
$\Omega$ is the angular velocity of the horizon, and $\Phi$ the electric potential at the horizon.  All quantities in the previous formula and in this article are given in geometric units where $G=1=c$ for translation into standard units see Appendix F of \cite{wald}. In these units the Planck constant $\hbar$ is an area $\hbar=\ell_p^2$ (with $\ell_p$ the Planck length).

Assuming the validity of the no-hair conjecture, the  previous relation can be directly derived by simple differentiation of the expression of the Kerr-Newman mass $M$ written as a function $A, J$, and $Q$, namely
\ba
&& M(A,J,Q)=4\pi\left(r_H^2+a^2\right)  \\
&& =4\pi
\left(\left(M+\sqrt{M^2-Q^2-J^2}\right)^2+\left(\frac{J}{M}\right)^2\right). \n 
\ea 
With some reasonable physical assumptions the process relating two stationary states can be  
described using the linearized Einstein's equations \cite{Wald:1995yp, Rignon-Bret:2023lyn}. Phase space derivations of the first law \cite{Wald:1993nt} 
are more rigorous, but can obscure some of the physical intuition that the linearized treatments offer.
The simplest is by far the one presented here which, like in standard thermodynamics, relies on the uniqueness of stationary states.

The first law expresses the conservation of energy 
in the context of an interaction with a black hole where the change in the mass of the system equals the work done by the perturbation (last two terms in \eqref{first})
plus the term $\kappa \delta A/(8\pi)$ which, we will now see, plays the role of heat. But for that one has to focus on the situation where quantum effects are no longer neglected. 

The first law (in its classical form) suggests the interpretation of the black hole area as an entropy measure of the black hole \cite{Bekenstein:1973ur}. This suggestion was first motivated by the discovery 
that the area of the black hole can only increase in every classical physical process. This follows from the classical positivity of energy that grants the 
focusing of the light-like geodesics in general relativity.  One has that
\be\label{arealaw}
\delta A\ge 0,
\ee
classically. This is the famous Hawking area theorem \cite{Hawking:1971tu} which, historically, was hinted upon by previous specific examples involving idealized particle exchanges with a black hole designed to 
extract energy from them (thermal machines)
in the Penrose mechanism \cite{Penrose:1971uk}, and the phenomenon of super-radiance \cite{Bardeen:1972fi} (see the definition by Christodoulou of the notion of irreducible mass \cite{Christodoulou:1970wf}).


Macroscopic Kerr-Newman black holes evaporate when quantum effects are taken into account: quantum fluctuations make them leak energy 
out to infinity in the form of black body radiation at the Hawking temperature 
\be\label{hh}
T=\frac{\kappa \ell_p^2 }{2\pi}.
\ee  
This follows from the calculation of the scattering of (test) matter fields prepared in the far past  in the vacuum
in a spacetime representing the dynamical process of gravitational collapse \cite{Hawking:1975vcx}. Renormalization of the energy momentum 
tensor in such quantum state leads to a result consistent with the interpretation that thermal energy at Hawking's temperature 
is radiated to infinity and extracted from the black hole \footnote{This can be shown unambiguously only in 2-dimensional models \cite{Fabbri:2005mw}. For more details
see \cite{Wald:1995yp}.}.  Calculations are valid for macroscopic black holes with areas $A\gg \ell_p^2$. 
The energy radiated by such black holes in their natural timescale (light transversing scale $\sqrt{A}$) is extremely small in comparison to their mass.
Therefore, even when they are not exactly in equilibrium (but rather slowly evaporating) they can be considered in a quasi-equilibrium state 
in the processes relevant for the discussion of the first law. In addition to the first and area law (the second law), black holes satisfy the zeroth law (Hawking temperature is the same in all directions) and the third law (extremal 
black holes, with $T=0$, cannot be formed from a finite sequence of physical processes) \cite{Bardeen:1973gs} .

\begin{figure}
\centerline{\hspace{0.5cm} \(\begin{array}{c}
\includegraphics[height=9cm]{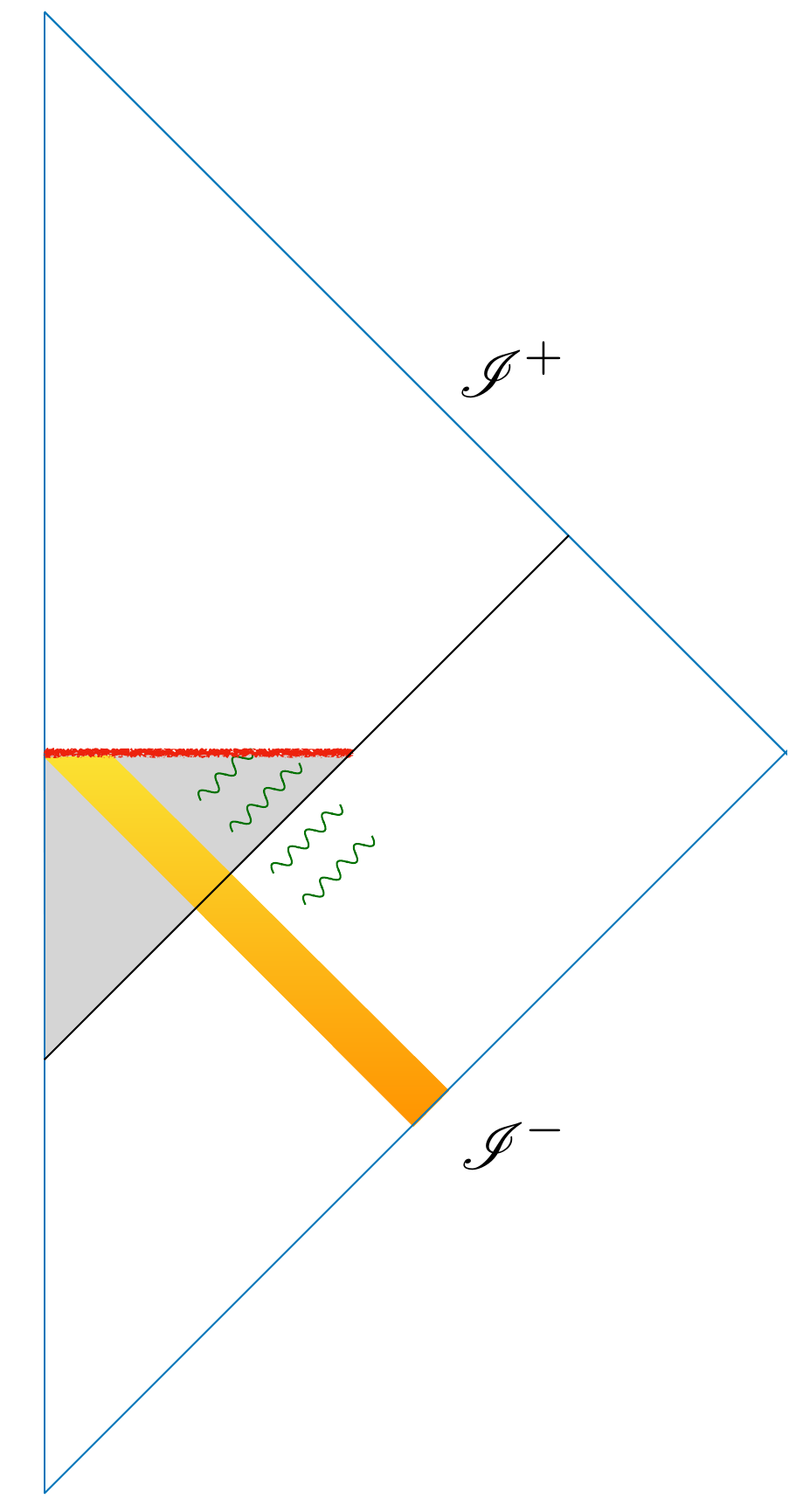} 
\end{array}\) } \caption{Penrose diagram representing the semiclassical representation of a black hole formed via gravitational collapse and subsequent evaporation. The black hole region is defined as the collection of events that are not connected to $\sI^+$ by causal curves 
avoiding the region of Planck scale curvature (classically referred to as the singularity). As in the classical case, the identification of the black hole region requires complete knowledge of the entire history of the system. Here we assumed that the black hole was formed by the collapse of null radiation sent from $\sI^-$. 
The light rays emitted at the end of evaporation (that extend from the horizon to $\sI^+$) represent the final pop of Hawking radiation \cite{Hawking:1974rv}. 
There are strong arguments against this representation: large quantum fluctuations invalidate the spacetime (mean field representation) in the future. See Figure \ref{cigar2}.}
\label{penroad}
\end{figure}

The discovery of Hawking radiation implies that black holes can be associated a temperature and thus the first law \eqref{first} becomes
\be\label{first-T}
\delta M=\overbrace{T \delta \left(\frac{A}{4\ell_p^2}\right)}^{\text{heat!}}+\underbrace{\Omega \delta J+\Phi \delta Q}_{\text{work terms}},
\ee 
with $T$ given by \eqref{hh}. The thermodynamic interpretation of the first law implies that the back hole should be associated an entropy 
\be
S_{\rm bh}=\frac{A}{4\ell_p^2}.
\ee
In the same spirit, area law \eqref{arealaw} would naturally be associated with the second law of thermodynamics. 
However, the positivity of energy that granted the focusing of the light-rays fails due to quantum effects: the hypothesis of the area theorem are violated and the horizon area 
may shrink due to the Hawking radiation. If the black hole is to be thought of as a thermal object then the second law has to include the entropy of the emitted radiation. 

Evidence for a version of the second law that includes the radiation can be found directly in 
the semiclassical description of the radiation process. Let us assume that the evaporation is slow (which is true for macroscopic black holes), and, for simplicity, that the background black hole is a vacuum and spherically symmetric one\footnote{A similar argument should hold for the rotating and charged black holes but it would demand further details about the radiation process.}. The evaporation process is well approximated by a sequence of Schwarzschild solutions with a time dependent mass $M$ that can be determined using conservation of energy.
Denoting $S_{\rm rad}$ the entropy content of the Hawking radiation received at infinity, we find that its rate of change is  
 given by \be \dot S_{\rm rad}=\frac{I_{E}}{T},
 \ee
where $I_E$ is the energy flux of radiation at infinity \footnote{This energy is characterized by the Stefan-Boltzman law   
\be\label{ii}
I_{E}=\sigma A T^4,
\ee
where $\sigma$ depends on the number of species involved and grey-body factors. 
As massive species will start contributing to the radiation only for suitably high temperatures, $\sigma$ will generally depend on $T$ and on our fundamental matter model \cite{Page:1976df}.
The following discussion is however independent of this issue.
}.
 Conservation of energy implies that
$\dot M=I_{E}$, and thus the previous equation can be simply expressed as \be \frac{dS_{\rm rad}}{dM}=\frac 1 T.\ee 
Using  that $T=\ell_p/(8\pi M)$ for the Schwarzschild black hole one can integrate the previous differential 
equation from an initial mass $M_0$ until complete evaporation. One finds 
\be\label{entro-pia}
\Delta S_{\rm rad }=\int_{M_0}^{0} \frac{8\pi}{\ell_p^2} M dM=\frac{A_0}{4\ell_p^2}=-\Delta S_{\rm bh}.
\ee
The previous equation tells us that the entropy content of the Hawking radiation matches $A_0/(4\ell_p^2)$ and reinforces the interpretation of the area dependent term in \eqref{first-T}
as a heat term. Furthermore, the result tells us that the total entropy of the system radiation+black-hole has remained constant in the evaporation process
which, due to the validity of the semiclassical approximation is, thus, adiabatic.
Deviations from reversibility can become important as the black hole shrinks towards the Planck scale where the semiclassical approximation fails (see Section \ref{crisis}). Furthermore, quasi-stationarity can be easily broken by feeding the black hole with matter from the outside in an uncontrolled manner after the initial black hole has formed.
Therefore, under general circumstances we expect that  
\be\label{gsl}
\Delta S_{\rm bh}+\Delta S_{\rm rad}\ge 0.
\ee  
Such a version of the second law is called the generalized second law (GSL) \cite{Bekenstein:1974ax}. It states that the total entropy of the universe  $S=S_{\rm bh}+S_{\rm out}$---where now $S_{\rm out}$ denotes the entropy of everything that is outside the black holes---can only increase in arbitrary physical processes relating equilibrium states, namely
\be\label{gfl}
\delta S\ge 0.
\ee
This is a new physical principle which integrates black hole entropy to the standard second law of thermodynamics. It declares that black holes contribute with $S_{\rm bh}=A/(4\ell_p^2)$ to 
the entropy in the standard thermodynamical sense. 

The GSL is postulated as a physical principle. As the standard second law one cannot actually prove its validity from first principles in absolutely general situations. This is because entropy 
increase in the universe is not a consequence of fundamental laws but rather a property of the very special nature of the `initial conditions' (the special state of the universe in the past) combined with our
inability (as macroscopic observers) to distinguish the microscopic degrees of freedom of the universe. The fundamental laws are essentially time reversal symmetric and any explanation of the arrow of time
implied by the second law is necessarily  not fundamental (at least in the present theoretical framework lacking any theory of initial conditions for the universe \cite{Penrose:1994de}).
Nevertheless, under certain simplifying circumstances, where the arrow of time is suitably coded in the boundary conditions defining the setting, where the nature of the degrees of freedom involved is postulated, and where an operational notion of entropy is introduced involving all the relevant degrees of freedom,  
it is sensible to try to prove the validity of the GSL.   

\section{Entanglement entropy as a measure of BH entropy}

The previous discussion makes clear that stationary black holes should be assigned an entropy given by a quarter of their area in Planck units. It is also clear that the first term 
in \eqref{first-T} must be interpreted as a `heat' contribution to the energy balance in processes involving perturbations of stationary black holes. We have reached this conclusion 
via a thermodynamical reasoning, similar to the one by Carnot, Joule, Clausius and others studying the energy balance in mechanical systems involving gases in the 19th century.
In the case of standard systems, the work of the founders of thermodynamics motivated the quest for a microscopic explanation of these thermodynamic laws that was later achieved via the statistical mechanics treatment 
of the (more fundamental) caracterization  of the constituents of matter. Similarly, the laws of black hole mechanics---discovered from the combination of general relativity and quantum field theory---call for a statistical description of black hole entropy in terms of more fundamental degrees of freedom, beyond those encoded in standard quantum field theory and classical gravity.

The immediate difficulty that one faces is that of finding a sensible working definition of entropy allowing for its computation in terms of (yet to be discovered) more fundamental degrees of freedom.
Here we recall that entropy is not a fundamental notion as it always involves some notion of coarse graining. This is at the heart of the thermodynamical description: the laws of thermodynamics
emerge in equilibrium and near equilibrium situations when the microscopic physics is suitably averaged out. In this respect,  a natural working definition is offered by the notion of entanglement entropy that we define below.

Entanglement entropy adds to the standard quantum mechanical (von Neumann) entropy of the degrees of freedom outside of the black hole, a contribution that quantifies the amount of entanglement between the exterior and the interior of the black holes. Only the black hole horizon remains in (marginal) causal contact with the outside, the microscopic contributions to entanglement across this limiting surface will be the ones accounting for the entropy of the black hole. From this perspective, the proportionality of black hole entropy with the area of the horizon is expected on general grounds from the microscopic structure of regular states in quantum field theory (and from the physics of black hole evaporation, see Section \ref{crisis}). The idea that entanglement entropy is a suitable notion capable of capturing the physics of systems containing black holes was first put forward in \cite{Bombelli:1986rw}. The definition that follows is phrased in the language of quantum mechanics. In quantum field theory (in the presence of local degrees of freedom) the expressions we write are formal due to the presence of ultraviolet (UV) divergences.    
As we will discuss in more detail below, divergencies come from contributions close to the surface separating the inside from the outside. When regulated by some cutoff, the divergent contribution to entanglement entropy is proportional to the area of the entanglement surface, which in applications to black holes coincides with the black hole horizon.  

The regularization dependence makes the entanglement entropy ambiguous; however, the fact that UV contributions from the entanglement surface are responsible of the area dependence is natural. It suggest that finding the correct formulation where divergencies disappear is the right track towards identifying the microscopic degrees of freedom responsible for the entropy of black holes. Such new physics is necessarily 
beyond the standard quantum field theoretic formulation. Note that UV divergences in the scattering theory approach of QFT are dealt with using renormalization theory. In gravity only the expectation value of 
the energy momentum tensor can be (in principle) renormalized and used in semiclassical Einstein's equations---$G_{ab}\!=\!8\pi\braket{T_{ab}}$---for insights into quantum gravity questions. 
However, there is no prescription to renormalize its more wildly UV divergent quantum fluctuations $\Delta_{abcd} \equiv \braket {T_{ab}T_{cd}}-\braket{T_{ab}}\braket{T_{cd}}$ and hence no
means to evaluate the validity of the semiclassical approach within standard QFT. Quantum gravity is needed to regularized the UV problem in QFT, the appearance of UV divergences in our working definition 
of black hole entropy is not surprising and in essence encouraging.
\begin{figure}[h!!!!!!!]
\centerline{\hspace{0.5cm} \(\begin{array}{c}
\includegraphics[height=8cm]{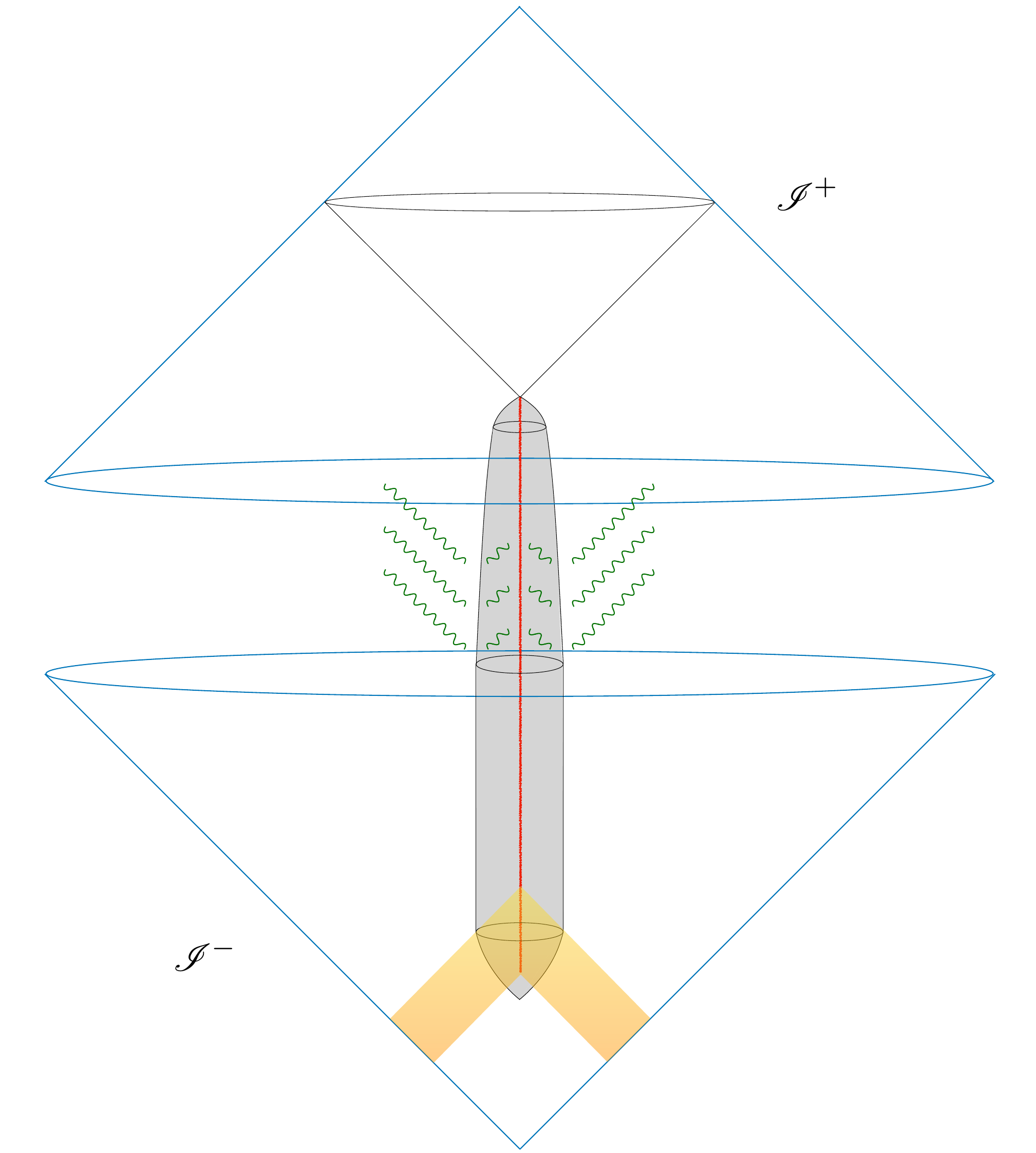} 
\end{array}\) } \caption{Adding an additional dimension to the representation of Figure \ref{penroad} is useful for intuition.
Here we show the semiclassical expectation for the complete history of black hole formation and evaporation. The last  moment of evaporation would correspond in this picture to a final flash of particles 
propagating along a lightcone in the (essentially) Minkowski spacetime left after complete evaporation. This picture is misleading in several aspects. An important improvement is shown in Figure \ref{cigar2}.}
\label{cigar1}
\end{figure}
\begin{figure}[h!!!!!!!]
\centerline{\hspace{0.5cm} \(\begin{array}{c}
\includegraphics[height=8cm]{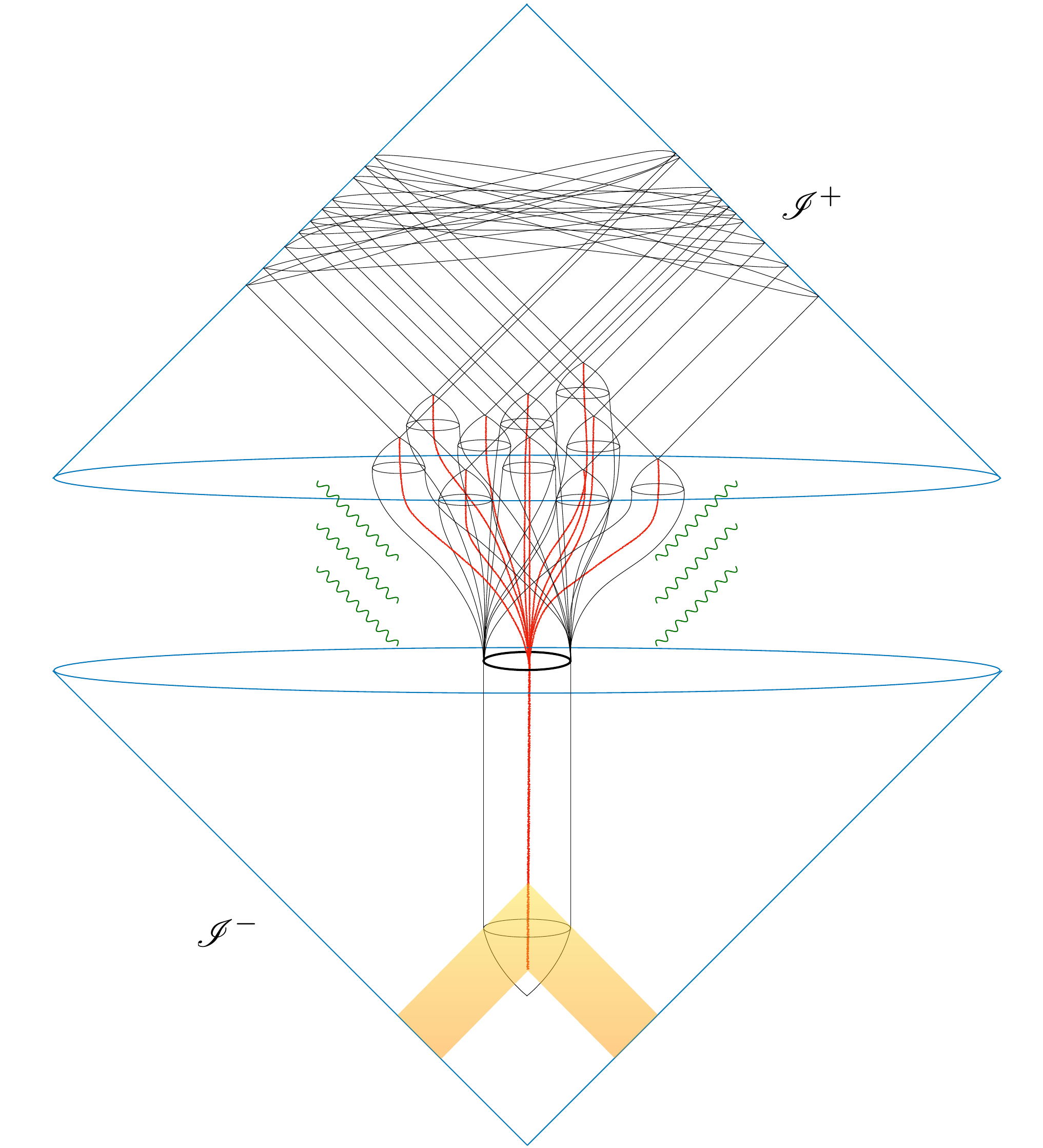} 
\end{array}\) } \caption{The back reaction of the particles emitted in the Hawking 
radiation on the black hole leads to a superposition of black hole geometries where the uncertainty in the black hole position is as large as $M^2$ in Planck units by the end of evaporation.
This shows that a fully quantum gravitational treatment is necessary well before the end of evaporation. The uncertainty in position is of the order of the Schwarzschild radius $M$ as soon as 
$\delta M/M>(m_p/M)^{2/3}$ (for a solar mass black hole this happens as soon as $\delta M_{\odot} \approx 10^{-25} M_{\odot}$ \cite{Page:1979tc}). Even when we do expect Page's estimate to be correct
(large fluctuations in the future) there is no quantitative way to do this calculation in the framework of standard QFT due to UV divergences in the formalism. An important aspect is that fluctuations are tamed 
and restricted to the UV neighbourhood of the black hole horizon when evolved backwards to the past, where a well defined semiclassical (coarse grained) horizon notion exists.}
\label{cigar2}
\end{figure}

The formal definition of entanglement entropy in QFT is the following: Consider a spacelike hyper-surface $\Sigma$ intersecting the black hole horizon at a two-dimensional surface and define the density matrix $\rho_{\rm out}$ as
\be
\rho_{\rm out}={\rm Tr}_{\rm in}[\rho],
\ee 
where $\rho$ is the density matrix representing the state of the universe, and the trace is taken over a complete basis of the Hilbert space representing local observables restricted to the inside of the black hole: the domain of dependence of the interior part of the Cauchy surface $\Sigma$. 
  
The state $\rho$ of the universe can be a pure state $\rho=\ket{\psi}\bra{\psi}$ or a mixed state (that would be the case for instance if one describes a statistical mixture; say a box with a gas in thermal equilibrium outside of the black hole, etc.). The entropy of the (outside) universe is defined by the von Newman entropy 
\be\label{tutin}
S=-{\rm Tr}[\rho_{\rm out} \log(\rho_{\rm out})].
\ee  
In the absence of black holes in the universe $\rho_{\rm out} =\rho$, the previous definition produces the quantum mechanical entropy of the universe which is non trivial  when  $\rho$ is mixed and vanishes for pure states $\rho=\ket{\psi}\bra{\psi}$. In the presence of black holes there are UV divergences from short wavelength modes close to the entanglement surface. If one introduces a length-scale cutoff $\epsilon$ then one finds \cite{Casini:2009sr}
\ba\label{entangly}
S&=&-{\rm Tr}[\rho_{\rm out} \log(\rho_{\rm out})]\n \\ &=& f_1 \frac{A}{\epsilon^2}+f_2 \log(A \epsilon^{-2})+S_{\rm out},
\ea  
where $f_1$ and $f_2$ are constants and $S_{\rm out}$ are finite contributions. The coefficient $f_1$ controls the leading area contribution to 
the entropy; however,  it depends on the 
regularization chosen (we will come back to this point below). The coefficient 
$f_2$ is a subleading correction to the entropy which is completely irrelevant from the thermodynamic perspective (such are tiny corrections in the macroscopic or thermodynamical limit).
Both $f_1$ and $f_2$ depend on the number of species of quantum fields involved \cite{Solodukhin:2011gn}.
Nevertheless,  unlike $f_1$, the coefficient $f_2$ is regularization independent in QFT. For this reason its calculation is a well defined problem that has received some attention 
in the literature for its mathematical physics interest \cite{Sen:2012dw}.

Let us consider the entanglement entropy in the following two instants of time (represented by Cauchy surfaces): first, $S(\Sigma_1)$, when a macroscopic black hole of area $A_0$ just formed and becomes stationary (this happens in a short time scale in comparison to the scale of evaporation), second, $S(\Sigma_2)$, when the black has almost completely evaporated via the emission of Hawking radiation.
In the first case we expect the entanglement entropy, \eqref{entangly}, to be dominated by the area (would be divergent) contribution
\be
S(\Sigma_1)\approx  f_1 \frac{A_0}{\epsilon^2},
\ee
while in the second case the entanglement entropy should be dominated  by the finite term $S_{\out}$  representing the entropy of the Hawking radiation emitted
during the evaporation. Namely, 
\be
S(\Sigma_2)\approx  S_{\rm rad}.
\ee
The change in the entropy is to be compared with our previous analysis leading to \eqref{entro-pia}. The comparison 
suggests that entanglement entropy is the correct measure of entropy characterizing the black hole and radiation system: 
the would be divergent term in \eqref{entangly} representing the entropy of the black hole. A mentioned, this contribution is ill defined in quantum field theory 
as it diverges due to the contribution of ultra short wave lengths near the horizon. On expects that a suitable definition 
based on the more refined understanding of the microscopic physics will cure these divergences. Moreover,  as the area of the black hole is controlled by Einsteins equations which become the Raychaudhuri equations when projected on the horizon, one expects that the number of species dependence of the would-be-divergent contribution to the entanglement entropy would also disappear when quantum gravitational effects are suitably taken into account.  Thus the expectation is that quantum gravity should regularize the would-be-divergent contribution and that the coefficient in front of the area will dynamically  emerge as the one fixed by semiclassical 
(or thermodynamical) considerations. Namely 
\be\label{lili}
 f_1 \frac{A}{\epsilon^2} \ \ \ \ \ \ \ \ {\begin{array} {ccc} {\longrightarrow}\\ \rm \van in\  quantum\\ \rm \van gravity\end{array}} \ \ \ \ \ \ \ \ \frac{A}{4\ell_p^2}.
\ee
The would be divergent term in the entanglement entropy appears as a reasonable candidate for the definition of black hole entropy.
It leads to a contribution proportional to the horizon area 
and it naturally incorporates the 
entropy contribution from matter degrees of freedom outside of the black hole. Divergences are expected to be regularized by quantum gravity, and  
the regulating mechanism is (in several models) linked to discreteness at the Planck scale. Available quantum gravity approaches
cannot (at their present level of developement) directly compute \eqref{entangly}. However, an observation 
from quantum field theory suggests an alternative view on the nature of the would-be-divergent term in \eqref{tutin} which provides 
the window through which present approaches can actually propose calculations of black hole entropy. 

In quantum field theory the microscopic structure of entanglement across a surface is universal for quantum states 
that admit a suitable renormalization of the expectation value of the energy momentum tensor. These states look like the 
vacuum state in the limit of short scales and are called Hadamard states. These states are compatible---in the semiclassical sense where one satisfies $G_{ab}=8\pi \braket{T_{ab}}$---with the description of gravity in terms of a regular (mean field) spacetime geometry and hence particularly interesting from the coarse grained perpective (behind the definition of black hole entropy we seek).
This is explicit in the vacuum case on Minkowski spacetime for suitable simple regions (basically spheres of arbitrary radia) and when one restricts the attention to conformally invariant fields.
It is also valid for more general fields if the entanglement surface is an infinite plane.
For such cases the reduced  vacuum state
$\rho_{\rm out}={\rm Tr}_{\rm in}(\ket{0}\bra{0})$ looks like a Gibbs state $\rho_{\rm out}=\exp({-2\pi H_{\rm m}})$ in terms of a natural Hamiltonian, the so-called modular Hamiltonian $H_{\rm m}$.
When expressed in terms of the (rescaled) Hamiltonian generating proper time evolution for an observer placed on an orbit of the boosts leaving invariant the entanglement surface, and hence
at constant proper simultaneity distance that we denote $\ell$, the reduced state is a thermal state with temperature $T=1/(2\pi\ell)$. In fact one can explicitly write the vacuum state in these cases as
\be\label{formy}
\ket{0}=\prod\limits_{\omega}\left[\sum\limits_n \exp{\left(-{\pi \ell} n\omega\right)} \ket{\omega, n}_{\rm in}\ket{\omega, n}_{\rm out}\right], 
\ee
where $\ket{\omega, n}_{\rm in/out}$ are Fock states with $n$ particles diagonalizing the Hamiltonian.
In the case of a plane the previous is the celebrated Unruh effect \cite{Unruh:1976db}.
The same decomposition of the Minkowski vacuum works in the case where the entanglement surface is a sphere (for conformally invariant fields) \cite{Perez:2023pcl}.
The previous equation shows that in the UV regime $\ell\to 0$ the temperature diverges 
and all excitations, as defined by an outside observer approaching the entanglement surface, become equally likely. 
The large number of out-states (maximally correlated with the in-states in \eqref{formy}) is the source of the divergent term in the entanglement entropy.   
The divergence would be regularized if the dimension of the Hilbert space of local excitation near the surface would be finite.
Something that is not true in quantum field theory, but could be realized in a quantum theory of gravity as we will explain below. 
Thus we see that the divergent term in \eqref{entangly} can be associated with microcanonical entropy in quantum gravity, namely 
\be\label{ruly}
 f_1 \frac{A}{\epsilon^2} \ \ \ \ \ \ \ \ {\begin{array} {ccc} {\longrightarrow}\\ \rm \van   Hadamard\end{array}} \ \ \ \ \ \ \ \ \log(N_A),
\ee
under the assumption that quantum gravitational effects render the dimension $N_A$ of the boundary Hilbert space
(compatible with a macroscopic area $A$) finite.  The argument linking the would be divergent contribution to entanglement entropy to microcanonical entropy 
is based on properties of the vacuum state in flat spacetime. However, as the divergent contribution at hand concerns the 
ultra local structure of the quantum state the previous discussion should apply to Hadamard states and for entanglement surfaces embedded 
in a low curvature environment, i.e., it should apply for macroscopic black holes.   

Let us also mention that regular notions built form entanglement entropy measures have led to interesting results in quantum field theory which are important for the present discussion and strengthen the idea of the relevance of the concept for black holes. The divergences 
we evoked here can be regularized when considering some related mathematical concepts such as that of 
relative entropy and mutual information. These concepts were used to prove some versions of the GSL 
in \cite{Wall:2011hj}, and Bousso's covariant entropy bounds \cite{Bousso:2014sda}, as well as in the analysis of certain Bekenstein bounds \cite{Casini:2008cr, Blanco:2013lea}.
The concept is also used as a guiding principle (as here) for insights about quantum gravity \cite{Bianchi:2012ev}.

\section{Computing entropy}

The simplest model is Wheeler's it-from-bit idea. If one postulates that the entanglement surface is made of discrete quanta of Planck area and degeneracy $g$.  Then 
$A/\ell_p^2$ area quanta are needed in order to produce a macroscopic area $A$. The dimension of the state space satisfying this constraint is $N_A=\exp(\log(g) {A/\ell_p^2})$
and the rule \eqref{ruly} gives the entropy 
\be
S_A=\log(g) \frac{A}{\ell_p^2}.
\ee
The previous heuristic calculation leads to a finite answer but it does not explain how the result is related to gravity in any way.
The idea of the discreteness of quantum geometry was at the times a wish vaguely motivated by general 
arguments in the context geometrodynamics \cite{Wheeler:1957mu}.
This wish became more solid with the discovery of the Ashtekar-Barbero \cite{Ashtekar:1986yd, BarberoG:1994eia} formulation of canonical quantum gravity where the kinematic phase space structure of
gravity---the one obtained when ignoring the constraints which are equivalent to some of the Einstein equations---is isomorphic to that of an $SU(2)$ Yang-Mills theory \cite{Ashtekar:2004eh}.
In terms of these variables the quantum uncertainty principle, derived from the classical Poisson brackets of gravity \cite{Freidel:2015gpa, Cattaneo:2016zsq}, implies the discreteness of the spectrum of area \cite{Rovelli:1994ge, Ashtekar:1996eg}.
Geometric operators like area and volume can be represented in a Hilbert space that is called kinematical \cite{Lewandowski:2005jk}. Quantum dynamics is expected to be described by quantum operators representing the classical constraints 
(for a recent review, perspective, and references see \cite{Ashtekar:2020xll}). Even when this remains an open question, the kinematical Hilbert space is though to be sufficiently large to accommodate the (infinitely many) local degrees of freedom of quantum field theory and gravity. However, it becomes finite dimensional (like in Wheeler's toy model) when restricted to the number of degrees of freedom of a surface constrained to have a given macroscopic area $A$.

The location of the horizon is, in principle, a complicated issue as one would need to know the entire evolution of the spacetime.
The assumption of equilibrium (stationarity) can simplify the problem at the classical level; however, the task remains very complicated at the quantum level (see Section \ref{crisis}).
The strategy found was to define the notion of a black hole horizon in equilibrium classically via suitable boundary conditions: the isolated horizon boundary condition \cite{Ashtekar:2000eq, Ashtekar:2004cn}.
The boundary condition carries a label $A$ corresponding to the classical area of the black hole. One quantizes gravity satisfying the isolated boundary condition with area parameter $A$ \footnote{This is analogous in a sense to the quantization of electromagnetism in a box of volume $V$ in order to describe the thermodynamics of photons in such a box.}. 
Due to the isolated horizon boundary condition, the degrees of freedom on the boundary admit a simple description in terms of Chern-Simons theory \cite{Engle:2009vc}. The quantum theory has an area operator whose spectrum is discrete: eigenstates of the area are given by a collection of punctures of the horizon carrying
spin quantum numbers. Namely,
\ba\label{area1}
&& A(S) |j_1,j_2\cdots\rangle=\n \\ && =\left(8\pi \gamma \ell_p^2  \sum_{p} \sqrt{j_p (j_p+1)}\right)\  |j_1,j_2\cdots\rangle,
\ea
where $\gamma$---the so-called Immirzi paremeter---is a dimensionless parameter labelling inequivalent kinematical 
quantum geometry Hilbert spaces.  There is a degeneracy $g_j=2j+1$ associated to the different configurations of the states with label $j$ at a given puncture.
The counting of the number of states $N_A$ is elementary in the case when $A\gg \ell_p^2 $ \cite{Ghosh:2008jc, Ghosh:2006ph}. If we define
$s_j$ the number of punctures of the horizon labelled by the spin $j$,  then the number of 
configurations is 
\be
N(\{s_j\})=\prod\limits_{j=\frac{1}{2}}^{\infty}\frac{n!}{s_j!}\,(2j+1)^{s_j},
\ee 
where $n\equiv\sum_j s_j$ is the total number of punctures and the multinomial factor comes from the assumption that the punctures are distinguishable. 
One defines the micro-canonical entropy as
$S=\log(n(\{\bar s_j\}))$ where $\bar s_j$ is the configuration maximizing the counting, namely 
\be\label{vary}
\delta \log(n(\{\bar s_j\}))+2\pi \gamma_0\delta C(\{\bar s_j\})=0
\ee
where $\gamma_0$ is a Lagrange multiplier of the constraint  
\ba\label{c1c2}
C(\{ s_j\})&=& \sum_j \sqrt{j(j+1)} s_j-\frac{A}{8\pi\gamma\ell_p^2}=0.
\ea
The solution of this variational equation is
\be\label{leading}
\frac{\bar s_j}{n}=(2j+1)\exp(- 2\pi\gamma_0 \sqrt{j(j+1)} ),  
\ee
which, when summing over $j$, leads to
\be\label{condition}
1=\sum_j (2j+1)\exp(- 2\pi \gamma_0 \sqrt{j(j+1)}).
\ee
The previous consistency condition fixes the numerical value of the Lagrangre multiplier $\gamma_0=0.274\cdots$, and 
evaluation on the maximizing configuration yields the following value for the black hole entropy
\be\label{92}
S=\frac{\gamma_0}{\gamma} \frac{A}{4\ell_p^2}.
\ee
When $A\gg \ell^2_p$ is not satisfied the counting is more involved and requires a more detailed analysis of the degeneracy of the area spectrum \eqref{area1}.
A series of papers have focuses on this aspect. Key representative examples are \cite{Domagala:2004jt, Agullo:2008yv, Agullo:2010zz, Agullo:2009eq} (for a review on these methods, results, and further references see \cite{BarberoG:2015xcq}).
It should be pointed out that as soon as black holes are not macroscopic one leaves the realm of quasistationary and enters the full dynamical quantum gravity regime 
where it is no longer clear  whether a meaningful notion of black hole entropy exists.

Three alternative interpretations of the previous result have been evoked in the literature. 
One is that consistency with black hole thermodynamics requires the fixing of the Immirzi parameter  so that $\gamma=\gamma_0$.
This view regards $\gamma$ as an ambiguity in the quantization that would be fixed from the requirement that the continuum limit leads to general relativity
and hence the usual laws of black hole thermodynamics. Another proposal, given in \cite{Jacobson:2007uj}, is that one should interpret the previous apparent mismatch as an indication of the renormalization of Newton's constant according to $G_N=\gamma G/\gamma_0$, where $G_N$ is the low energy Newton's constant and $G$ the microscopic one entering the definition of the fundamental scale $\ell_p$ and the spectrum of the area \eqref{area1}.
The previous counting misses the contribution of matter degrees of freedom near the horizon (only geometry plays a role in the previous counting).
The third view is that the necessary inclusion of matter (see Section \ref{crisis}) would affect the degeneracy factors $g_j$ in the previous counting. Following physical ideas in the context of the brick-wall paradigm \cite{tHooft:1984kcu} suggests that this could eliminate the Immirzi parameter dependence in the entropy producing a result in agreement with the semiclassical expectations \cite{Ghosh:2013iwa}. 

Another approach for calculating black hole entropy is the one proposed by the GFT formulation of quantum gravity \cite{Oriti:2006se}. In such framework spherically symmetric horizon geometries are described in terms of a condensate of geometric quanta
closely related to those of loop quantum gravity. The averaging effect of the condensate state is claimed to erase the dependence on $\gamma$ and to produce an entropy in agreement with the Bekenstein-Hawking semiclassical value \cite{Oriti:2015qva,Oriti:2015rwa, Oriti:2018qty}.

The question of which of the previous views, if any, is the correct one remains unsettled. Perhaps this is due to the difficulty in evaluating the contribution to the entropy of the true dynamical degrees of freedom close to the black hole horizon.  The isolated horizon boundary condition is a nice simplification that allows for a precise calculation. However, it might restrict too severely the freedom in view of physical expectations that we briefly evoke in the next section.  

\section{The semiclassical crisis}\label{crisis}

The previous discussion presupposes that one can define the location of the black hole horizon. In the classical $\hbar=0$ case (where black holes do not evaporate) one defines de black hole horizon as the frontier of the past of future null infinity $\sI^+$  (see \cite{wald} for details on asymptotic flatness and Penrose's conformal compactification). The teleological character of the horizon, which makes it a causal frontier, is important for instance in the proof of Hawking area theorem and several other theorems concerning black holes in general relativity. In this definition the horizon separates the outside (what can be seen from infinity) from the rest. That is the object behaving thermodynamically. 

Assuming a suitable form of strong cosmic censor holds \cite{Penrose:1999vj} and the singularity inside is a spacelike frontier to the future of all observers inside, then one could define the black hole region as the past domain of dependence of the singularity. Such a definition is not practical in the classical theory because of the complications associated with identifying the singularity especially when the interior geometry of stationary black hole solutions is known not to represent the geometry of the interior of real black holes formed via gravitational collapse \cite{wald}. However, such definition remains available in the quantum gravity context where dealing with the interior singularity is, in any case, necessary (see also \cite{Perez:2022jlm} for a related definition). Thus, when we turn on $\hbar$ the black hole evaporates and eventually disappears that second perspective still makes sense if one defines the singularity as the region of Planckian curvature and Planck scale fluctuations. In some approaches the singularity becomes naked (a strong quantum gravity region visible from infinity) with the classical black hole interior in causal contact with the outside at the end of evaporation (see Figure \ref{penroad} and Figure \ref{cigar1}) yet this does not conflict our definition. This is the definition used in Figures \ref{penroad} and \ref{cigar1}.

As pointed out by Page \cite{Page:1979tc} the physics of Hawking radiation however implies the necessity of important modifications in the semiclassical picture presented in Figure \ref{cigar1}. Using momentum conservation one can estimate the recoil of the black hole due to the emission of the individual Hawking particles. This implies that after some time (and while the black hole is still macroscopic) the black hole and radiation emitted would be in a quantum superposition where the position of the black hole shows extremely large quantum fluctuations. As the evaporation time is very long---$M^3$ in Planck units---the fluctuations in position grow to the astronomical scale of the order of $M^2$ . This means that the picture of a semiclassical object (classical spacetime) emitting radiation until the end of evaporation is inappropriate. Rather, the image that emerges is that of a superposition of essentially flat (yet all inequivalent) quantum geometries by the end of evaporation. Each and every one of these geometries is filled with matter quanta corresponding to the history of emission that led to that particular final location of the last explosive evaporation event \cite{Hawking:1974rv}. The representation in Figure \ref{cigar1} should be abandoned in favour of the cartoon in Figure \ref{cigar2} where we have superimposed the various histories and the traces of their explosive end at a common $\sI^+$. Notice that this effect already shows that we have underestimated the radiation entropy at the end of evaporation in our semiclassical account and that the inequality in \eqref{gsl} is justified. 

A good news is that, as Figure \ref{cigar2} emphasizes, the large fluctuations of the black hole region in the far future become negligible towards the past (where the black hole is macroscopic and young)
and confined to the UV structure of the black hole region (matter and geometry) near the horizon. In the past a coarse grained notion of black horizon still makes sense as well as the computation of its entropy.
The Planckian differences in the location, shape, and matter configurations of the horizon and its immediate UV atmosphere in the past, obtained from the backward evolution of the future histories, define the micro-states responsible for black hole entropy. These are the ones to be counted: these are no internal states of the black hole but rather the many micro-states by which the states representing the black hole differ as seen from the outside.

\section{Implications for unitarity}

The first intuition is that anything that crosses the black hole horizon becomes causally disconnected from outside observers. This appears to be true for the matter and gravitational degrees of freedom that formed the black hole in the first place or were absorbed later in its future evolution. It is also true (within the QFT framework) for the maximally correlated inside-partners of the Hawking modes radiated to infinity \footnote{Particles created by a gravitational field are produced in perfectly correlated pairs \cite{Wald:1975kc}.
In the case of Hawking radiation one of these particles goes out to infinity and its partner falls to the inside singularity}. In both cases the relevant degrees of freedom evolve in the highly non stationary environment of the black hole interior and fall to the singularity region where strong quantum gravity effects are unavoidable. The expected causal structure inside of the black hole tells us that the fate of what formed the black hole and later led to its slow evaporation is written in the region where only a quantum gravity theory can describe the physics. 

Now, if black holes completely evaporate, due to Hawking radiation, the inside with its quantum gravity region (that we call the singularity) and all the information fell would seem to disappear at the end of evaporation. The situation is simpler to analyse if one focusses for a moment on the semiclassical description of the Hawking radiation. In the usual account it would seem that one starts with vacuum state in the far past ($\sI^-$ in Figure \ref{penroad}) and evolves into a state that is mixed containing the Hawking thermal radiation slowly emitted by the black hole. Such evolution is forbidden in a unitary theory revealing that either information is lost and quantum gravity must cope with the breaking of unitarity in quantum theory, or that the information remains in some degrees of freedom that make the final state pure (despite apparences). This is the Hawking information puzzle \cite{Hawking:1976ra}.

The debate has lasted for almost 50 years. The absence of a definite answer resides in the difficulty of analysing the strong quantum gravity regime near the singularity. 
After all it is clear that this region must play a central role in the understanding of the physics involved: from the semiclassical perpective evolution is expected to be 
perfectly unitary for the quantum fields involved in Hawking radiation between Cauchy surfaces that (necessarily) approach the singularity region with inner volumes that become astronomical 
inside the black hole \cite{Christodoulou:2016tuu, Wheeler:1964qna}. Even when non-local aspects of quantum gravity are not to be neglected (see below) this insight from QFT suggest strongly 
that the dynamics near the singularity is also a key aspect of the problem.

In the absence of a reliable theory capable of dealing with the microscopic physics near the singularity 
people have tried to find low energy (macroscopic) channels for information to be retrieved. As mentioned in the previous section, effects of the radiated quanta 
that are not encoded in $\braket{T_{ab}}$ but in its fluctuations lead to a final state where there a huge uncertainty in the black hole position of the order 
of $M^2$ in Planck units \cite{Page:1979tc}. This shows that the perfectly thermal spectrum at infinity would be modified by fluctuations that are not taken into account.
Could the apparently lost information be written in these deviations from thermality? 
Even when this aspect is certainly important---as argued it completely invalidates many of the representations we make of black hole evaporation---its effects do not seem enough to resolve the puzzle of information
(see \cite{Almheiri:2020cfm} and \cite{Flanagan:2021ojq} and references therein for different ideas on such view).    

Trying to answer the question in the previous paragraph one can push the puzzle to an intuitive extreme where one does not deal with correlations in Hawking radiation but rather with macroscopic information.  
Assume that Alice fell with a book inside of a black hole (for a sufficiently massive black hole) one would need to show that  fluctuations of the energy momentum tensor, accessible to Bob outside of the black hole,
can be used to recover the information in Alice's book. But information cannot be doubled in quantum mechanics, so would these fluctuations erase the pages in Alices book as she approaches the singularity?
There are many proposals exploring how information could actually come out with the Hawking radiation and thus save unitarity. 
All of these fail in giving a precise account of what actually happens once the relevant degrees of freedom interact with the Planckian ones at the singularity.
None of these describes how the story of Alice continues as she falls into the quantum gravity region near the singularity.

We have seen that discreteness at the Planck scale can regularize the divergences in the definition of the entanglement entropy.
We argued that the divergent contribution in quantum field theory is related (as a microcanonical entropy) to the number of microscopic states
associated with a coarse grained area $A$. The interpretation that follows is that for every smooth macroscopic field theoretical description of gravity and matter 
there is a large number of microscopic discrete degeneracy. The continuum physics would be emergent from the discrete with the microscopic granularity regulating the
UV divergences that plague quantum field theory. If its role is central in defining black hole entropy, it should be central too in describing
the fate of information. Alice and her book (as well as the Hawking partners of the black hole radiation) fall into the strong field Planckian regime
and (both infinitely blue shifted by tidal effects) interact directly with the microscopic degrees of freedom. These are overwhelmingly more than those of the continuum as already the black hole entropy calculation shows.
Alice's story continues but now is written in the fundamental building blocks at the Planck scale.  On entropic grounds the information coded in such micro-structure of a granular Planckian physics cannot be retrieved in terms of low energy information in the future when the black hole has evaporated away. This would require a conspiracy that we have learnt never to happen as it is written in the foundations of our statistical mechanical understanding of the macroscopic world and thermodynamics. Thus information rescuing unitarity would remain written in a pure superposition of inaccessible Planckian defects in the fabric of a flat (coarse grained) mean-field spacetime entangled with the particles in the Hawking radiation emitted during the long evaporation history of the black hole \cite{Perez:2014xca, Amadei:2019wjp, Perez:2023ugg}.

It is however important to point out that the distinction between the inside and the outside  
of a black hole (or any two regions in spacetime) is blurred in quantum theory and that the issue becomes even more problematic in quantum gravity. In quantum field theory this is 
due to the non local correlations present in suitable Hadamard states that is among other things  
responsable of the ultraviolet divergences in the entanglement entropy. 
In quantum gravity this non locality becomes even stronger due to the presence of the gravitational constraints implying diffeomorphism invariant 
observables are non local \cite{Jacobson:2012ubm, Jacobson:2019gnm}. This non locality combined with the one of quantum mechanics cannot be neglected in discussions about the fate of information in black hole evaporation. 
Even when Alice's book would be destroyed at the singularity information in the book is non locally distributed (in a quantum mechanical sense) in the state of the universe and perhaps retrievable 
across causal barriers such as the black hole horizon \cite{Raju:2021lwh, Chowdhury:2021nxw, Laddha:2020kvp}. 
The previous discussion shows that non-locality and the Planckian discreteness must be   
part of a single consistent description of the unitarity question. I believe that this question is crucial and  
 deeply related to the fundamental structure of quantum gravity.

  \section*{Acknowledgements}
I am grateful for exchanges with D. Page and L. Freidel.
This work was made possible through the support of 
the ID\#62312  grant from the John Templeton Foundation, as part of the ``The Quantum Information Structure of Spacetime (QISS)'' Project (\href{qiss.fr}{qiss.fr}).

\providecommand{\href}[2]{#2}\begingroup\raggedright\endgroup


\end{document}